# Optical Refrigeration for Ultra-Efficient Photovoltaics

Assaf Manor[1], Leopoldo L. Martin[2] and Carmel Rotschild[1,2]

*[1] Russell Berrie Nanotechnology Institute, Technion − Israel Institute of Technology, Haifa 32000, Israel*
*[2] Department of Mechanical Engineering, Technion − Israel Institute of Technology, Haifa 32000, Israel*
Email: carmelr@.technion.ac.il

**Abstract**

Improving the conversion efficiency of solar energy to electricity is most important to mankind. For single-junction photovoltaic solar-cells, the Shockley-Queisser thermodynamic efficiency limit is extensively due to the heat dissipation, inherently accompanying the quantum process of electro-chemical potential generation. Concepts such as solar thermo-photovoltaics and thermo-photonics, have been suggested to harness this wasted heat, yet efficiencies exceeding the Shockley-Queisser limit have not been demonstrated due to the challenge of operating at high temperatures. Here, we present a highly efficient converter based on endothermic photoluminescence, which operates at relative low temperatures. The thermally induced blue-shifted photoluminescence of a low-bandgap absorber is coupled to a high-bandgap photovoltaic cell. The high absorber's photo-current and the high cell's voltage results in 69% maximal theoretical conversion efficiencies. We experimentally demonstrate tenfold thermal-enhancement of useful radiation for the high-bandgap cell and 107% enhancement in average photon energy. This paves the way for introducing disruptive-innovation in photovoltaics.

**One Sentence Summary:**

We experimentally demonstrate endothermic emission that increases solar cells efficiency by harvesting thermal losses. Efficiency of 69% is theoretically predicted.

**Main Text:**

The conversion-efficiency of single-junction photovoltaic (PV) cells has increased greatly over the last few decades (*1–6*), approaching its fundamental Shockley-Queisser (SQ) limit, of about 30% without solar concentration, and 40% under highest solar concentration (*7*). For low-bandgap PVs, the main loss mechanism is the generation of wasted heat due to the difference between the photon energy and the generated energy at each conversion event, which is set by the PV's bandgap. Many concepts have been suggested to harness this wasted heat for efficiency enhancement. These include solar thermo-photovoltaics (STPV) (*8–10*) and thermo-photonics (TPX) (*11, 12*) where an intermediate absorber is heated by sunlight, and emits radiation with a greater photon rate towards the solar cell. Similar aim for thermionic energy conversion was suggested in the form of Photon Enhanced Thermionic Emitter (PETE) (*13*). Although theoretically such concepts predict high efficiencies, to date, none of them have been experimentally proven to overcome the SQ limit. This is mainly due to the very high operating temperatures, which significantly challenge electro-optical devices. A preferable thermodynamic concept for harvesting wasted energy is by enhancing the extracted energy quanta per photon, and enhancing the induced PV voltage, while conserving the photon current. Such concept benefits from low entropy generation, which results in comparably low temperatures. Here, we present endothermic photoluminescence (PL) for high-efficiency single-junction PV. Specifically, a thermally-insulated low-bandgap photoluminescent absorber, absorbs the solar spectrum above its bandgap and emits towards a higher bandgap PV cell, maintained at room temperature. Similarly to optical refrigeration (*14–16*), each emitted photon is thermally blue shifted and matched to a higher bandgap PV cell, which converts the excessive thermal energy to enhanced electricity. Such arrangement benefits from both the high photon current of the low-bandgap absorber, and the high voltage of the high-bandgap PV, which yields conversion efficiencies approaching 68%, at practical operating temperatures. In addition, the system's optical and electrical components are de-coupled, which simplifies its realization.

Figure 1A shows our conceptual design. Solar radiation impinges on a thermally-insulated photoluminescent absorber, where its "hot" PL is coupled to a PV cell maintained at room temperature. In order to minimize radiation losses, a hemi-ellipsoidal reflective dome recycles photons by reflecting back emission at angles larger than the sun's solid-angle, $\Omega_1$. (*7, 17–19*). In addition, the PV's back-reflector which is usually designed to recycle the PV's inherent emission (*20, 1*), is used to reflect sub-bandgap photons back to the absorber. The "hot" PL spectrum is thermally modified compared to the room temperature PL as can be seen qualitatively in figure 1B. Figure 1C shows the absorber and PV energy levels, the endothermic PL and the current flow, as well as the reflected photons from the PV. The absorber's thermodynamics is governed by two conservation rules: **i.** Conservation of energy. **ii.** Conservation of rates, also known as the detailed-balance principle (*7*), which takes into account the absorbed solar photons and spontaneous emission of both absorber and PV. At the PV, only conservation of rate is applied between the incoming PL, the PV's emitted light and the extracted electric current. Energy conservation does not apply, because heat is extracted from the PV to keep it at room temperature. Figure 1D depicts the different factors in the detailed-balance (dotted arrows), with additional heat extraction (solid red arrow).

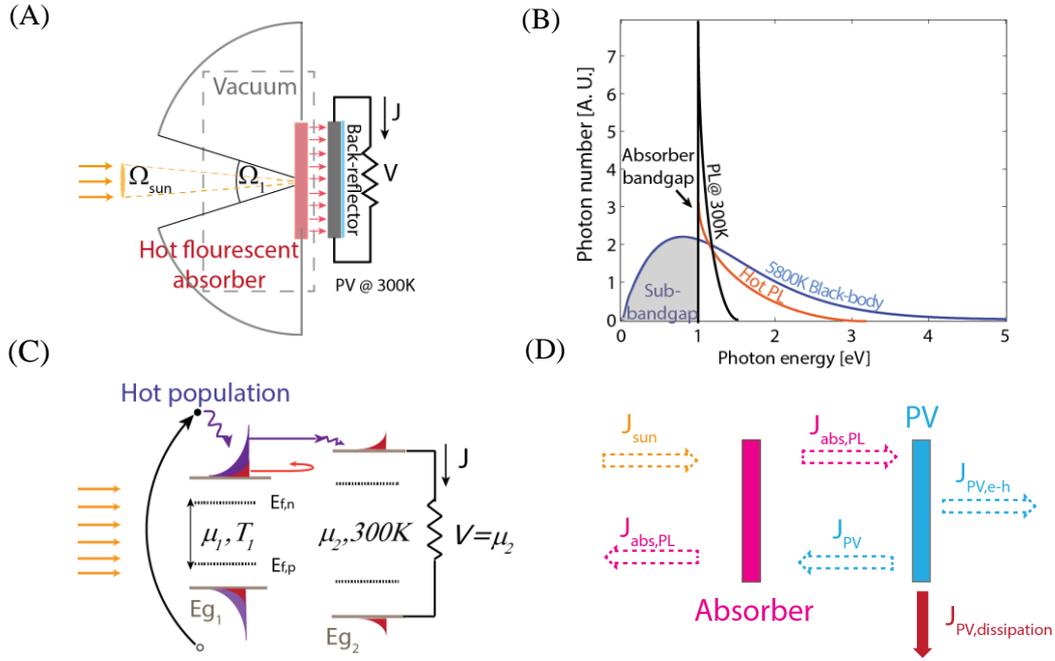

**Fig. 1.** (A) The device's scheme (B) Illustration of the endothermic PL Vs. the solar spectrum and room temperature PL (C) The absorber/cell energy levels and current flow (D) Factors in the detailed balance of rates (dotted arrows) and heat extraction (solid red arrow)

Within the detailed-balance, the master equation suiting the spontaneous emission rate of a bandgap material is the integration through all frequencies of the generalized Planck law, derived by Wurfel (*21, 22*):

$$R(\hbar\omega, T, \mu) = \varepsilon(\hbar\omega) \cdot \frac{(\hbar\omega)^2}{4\pi^2 \hbar^3 c^2} \frac{1}{e^{\frac{\hbar\omega - \mu}{KT}} - 1} \cong \varepsilon(\hbar\omega) \cdot R_0(\hbar\omega, T) \cdot e^{\frac{\mu}{K_b T}} \tag{1}$$

Where $R$ is the emitted photon rate (photons per second per unit area), $T$ is the temperature, $\mu$ is the chemical-potential, $\varepsilon$ is the emissivity and $K_b$ is Boltzmann's constant. The approximation holds as long as the material is far from lasing conditions. The chemical-potential ($\mu>0$) defines the excited states population deviation from thermal equilibrium. At $\mu=0$, the PL radiation reduces to the thermal emission rate ($R_0$). The energy current is simply calculated by multiplying $R$ by the energy quanta per photon, $\hbar\omega$.

Under the assumption of full absorption above the bandgap and unity quantum-efficiency (QE) for both the emitter and the PV, The detailed-balance is uniquely solved by fixing the only free parameter in the system; the PV voltage. This defines the absorber's thermodynamic properties, $T$ and $\mu$, as well as the different currents. A rigorous analysis of the detailed-balance is given at the method section.

We first use the detailed-balance to optimize $\Omega_1$ (see methods section), and find that the maximal efficiency is when the photon-recycling angle is maximal. In this case, the acceptance solid angle of the sun equals its viewing angle without any use of concentrating lens. ($\Omega_1=\Omega_{sun}=6.94\cdot10^{-5}$ *Srad*).

Next, solving the detailed balance for various voltages yields the system's I-V curve. Figure 2A depicts three I-V curves calculated for $Eg_1=0.7\ eV$, while $Eg_2$ varies between *0.7 eV, 1.1 eV*, and *1.5 eV*. To be more applicative, the photon-recycling ratio is set to a moderate factor, equivalent to solar concentration of 5000 suns (see method section). The I-V curve for $Eg_1=Eg_2=0.7\ eV$ (red line) shows a remarkable, "double humped" feature, and includes a "hot" and a "cold" Maximal Power Points (MPP). While at the "hot" MPP enhanced current leads to 53.6% efficiency, the "cold" MPP remains at the SQ efficiency limit of 33%. When $Eg_2$ is increased, the current enhancement vanishes and replaced by voltage enhancement, which for $Eg_2=1.5eV$ sets an efficiency of 65.6% at the cold MPP (blue curve).

Figure 2B shows the change in absorber's temperature and chemical-potential as a function of the operating voltage at $Eg_1=Eg_2=0.7$ eV. At low operating voltages, the chemical-potential remains zero, and the absorber is a pure black-body ($\mu=0$), characterized only by its temperature and its step-function emissivity at its bandgap. These operating conditions are similar to the solar-thermal PV concept suggested by Wurfel (*8*).

A thermodynamic transition occurs at V=0.55V, where the temperature drops sharply while the chemical-potential increases. Above the transition voltage, the dominant effect is the PV's spontaneous emission, which grows exponentially with the voltage (*7*). The PV cell is kept at 300K, therefore its emission is spectrally narrow compared to the endothermic PL (inset in Figure 2B). The recycling of the PV emission has the effect of optical refrigeration that drops the temperature and increases the absorber's chemical-potential (Figure 2B). Intuitively, the QE conserves the emitted photon rates (*21*), hence the only way to balance both the absorbed energy and photon rate is when each emitted photon also carries thermal energy, making these photons accessible to higher bandgap PV.

By raising $Eg_2$ above $Eg_1$ (Figure 1C, Figure 2A blue and green curves), **the extracted energy at the cold MPP benefits from both the high photon-current of the low-bandgap absorber, and the high voltage of the high-bandgap PV**. This yields conversion efficiencies as high as 65.6%, at practical operating temperatures of up to 1300K. We note that under full photon-recycling, due to maximal optical-cooling, the efficiency improves to 69% and temperature reduced to 1180K.

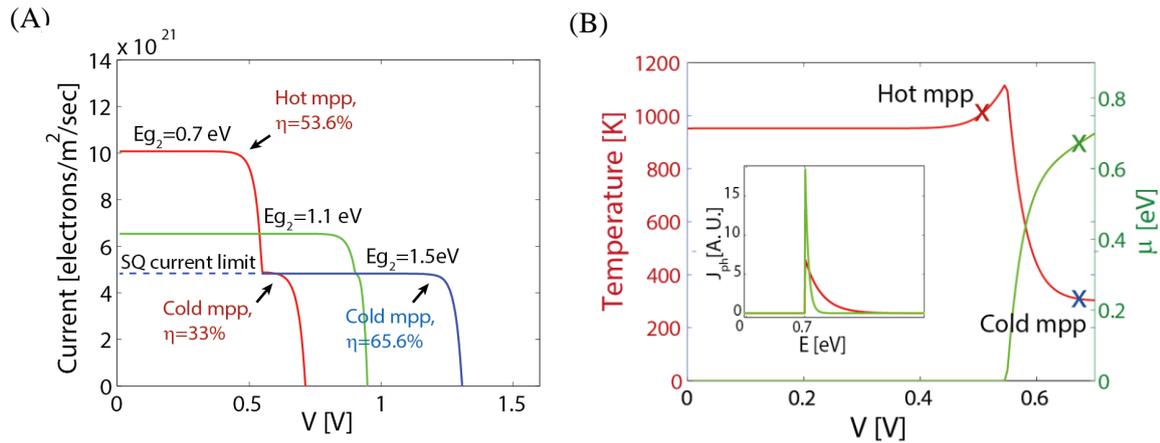

**Fig. 2.** (A) I-V curves corresponding to a *0.7 eV* absorber and PV bandgap values of (*0.7 eV, 1.1 eV* and *1.5 eV*). (B) The temperature and chemical-potential dependence on the PV cell voltage for $Eg_1= Eg_2=0.7$ *eV*. Inset: comparison of the hot absorber and the cold cell spectra.

We continue with an experimental demonstration of the cross-over from thermal to PL regimes and the accompanied drastic change in operating temperatures, as well as the thermal enhancement of the accessible photons for high-bandgap PV. Experimentally, we need to control both the photonic and thermal loads on a PL material, which conserves its QE at high temperatures. Rare-earth ions such as Neodymium and Ytterbium are excellent choice of materials as they conserve their high QE at extremely high temperatures (*23*). This is in sharp contrast to solid-state semiconductors where temperature dependent non-radiative recombination reduces QE (*24*). It is important to clarify that simply measuring PL under different temperatures using an oven (fixed temperature bath) does not demonstrate the detailed-balance, because in our system the temperature is a product of the balance between the thermal and photonic currents.

For this reason, we study the fluorescence of Neodymium ($Nd^{+3}$) doped Silica fiber tip under 532 nm photonic excitation. The thermal current is supplied by a $CO_2$ laser operating at 10.6μ wavelength. At such wavelength the photons are efficiently absorbed by the Silica (*25*), and converted to a constant heat-flow. The experimental setup is sketched in figure 3A. The PL is measured by a calibrated fiber coupled spectrometer. The PL excitation at *532nm* is kept constant at *1mW*, while the $CO_2$ power varies between 0 to *150mW*. The results are shown in figure 3B, while figure 3C shows the corresponding total number of emitted photons at shorter wavelength than 1μm. Evidently, as we increase the thermal load, the PL exhibits a blue-shift evolution. This is shown by the reduction of the low photon energy, 900nm emission peak, and enhancement of the high photon energy, 820nm emission peak (Figure 3B), while photon number is conserved (green solid line at Figure 3C). **We**

**measure the average gain in photon's energy to be ~0.1eV, which corresponds to 107% power enhancement relative to the cold PL**.

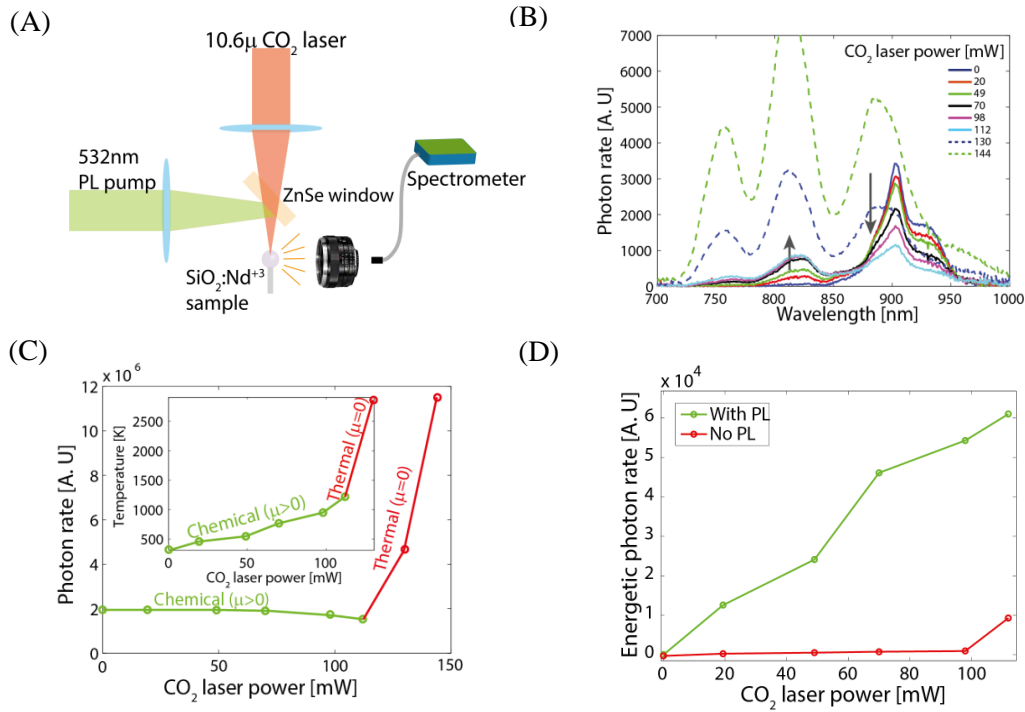

**Fig. 3.** Experimental setup (A) PL spectra evolution with $CO_2$ power increment (B) photon rate and temperature dependence on $CO_2$ power (C) comparison of the energetic photon number, below 850nm (D)

This trend continues until reaching the transition point at $CO_2$ power levels of 110 $mW$. As we increase the heat current even further, the power spectrum and the number of emitted photons increase sharply at all wavelengths (dotted lines in Figure 3B and red line in 3C). In addition, we monitor the operating temperature by Fluorescence Intensity Ratio Thermometry (FIR) (*26*) (Figure 3C, inset). As one can see, the temperature rises monotonically with increasing the heat load. Above the transition point there is a dramatic increase in the temperature slope. (See method section for the detailed FIR description). **This is a clear evidence of the transition from the PL ($\mu$>0) regime to thermal regime ($\mu$=0) where temperature rises sharply and conservation of photon number does not apply.** Since our vision is to couple the emission to a high bandgap PV, measuring the average photon energy enhancement is insufficient. This is because the photon-recycling in the full configuration would reflect low energy photons, at wavelength longer then 850nm, back to the absorber. For this reason, it is important to compare the number of accessible energetic photons, at wavelength shorter than 850nm, between the PL and the thermal regimes. This is done by turning "on" and "off" the PL pump (at 532 nm) under constant thermal current, operating below the transition point. Figure 3D presents this experimental comparison. **When the PL**

**pump is "on", the energetic photons rate is higher than at "off", by an order of magnitude.** This indicates on the potential improved conversion efficiency when using a high-bandgap PV to harvest the thermally enhanced PL over pure thermal emission at similar absorbed energy rates.

To conclude, a novel concept for efficient conversion of solar energy to electricity based on endothermic PL and photon-recycling has been presented. An experimental demonstration shows that the emission rate of energetic photons, accessible to high-bandgap PV, is enhanced by an order of magnitude and the operating temperature reduced sharply. This paves the way for introducing disruptive-innovation in photovoltaics. Future research will focus on tailoring the absorption spectrum of the PL material to the solar spectrum, and its endothermic emission to state of the art photovoltaic cells.

**References and Notes**

1. B. M. Kayes *et al.*, in *2011 37th IEEE Photovoltaic Specialists Conference (PVSC)*, (2011), pp. 000004–000008.

2. I. Repins *et al.*, *Prog. Photovolt. Res. Appl.* **16**, 235–239 (2008).

3. J. Zhao, A. Wang, M. A. Green, F. Ferrazza, *Appl. Phys. Lett.* **73**, 1991–1993 (1998).

4. M. Liu, M. B. Johnston, H. J. Snaith, *Nature* **501**, 395–398 (2013).

5. P. Peumans, S. Uchida, S. R. Forrest, *Nature* **425**, 158–162 (2003).

6. S. Gélinas *et al.*, *Science* **343**, 512–516 (2014).

7. W. Shockley, H. J. Queisser, *J. Appl. Phys.* **32**, 510 (1961).

8. P. Wurfel, W. Ruppel, *IEEE Trans. Electron Devices* **27**, 745–750 (1980).

9. R. M. Swanson, *IEEE Proc.* **67**, 446 (1979).

10. A. Lenert *et al.*, *Nat. Nanotechnol.* **9**, 126–130 (2014).

11. N. P. Harder, M. A. Green, *Semicond. Sci. Technol.* **18**, S270 (2003).

12. I. Tobias, A. Luque, *IEEE Trans. Electron Devices* **49**, 2024–2030 (2002).

13. J. W. Schwede *et al.*, *Nat. Mater.* **9**, 762–767 (2010).

14. D. V. Seletskiy *et al.*, *Nat. Photonics* **4**, 161–164 (2010).

15. M. Sheik-Bahae, R. I. Epstein, *Phys. Rev. Lett.* **92**, 247403 (2004).

16. M. A. Weinstein, *J. Opt. Soc. Am.* **50**, 597 (1960).

17. A. Polman, H. A. Atwater, *Nat. Mater.* **11**, 174–177 (2012).

18. E. D. Kosten, J. H. Atwater, J. Parsons, A. Polman, H. A. Atwater, *Light Sci. Appl.* **2**, e45 (2013).

19. A. Braun, E. A. Katz, D. Feuermann, B. M. Kayes, J. M. Gordon, *Energy Environ. Sci.* **6**, 1499–1503 (2013).

20. E. Yablonovitch, O. D. Miller, S. R. Kurtz, in *2012 38th IEEE Photovoltaic Specialists Conference (PVSC)*, (2012), pp. 001556–001559.

21. P. Wurfel, *J. Phys. C Solid State Phys.* **15**, 3967–3985 (1982).

22. F. Herrmann, P. Würfel, *Am. J. Phys.* **73**, 717 (2005).

23. G. Torsello *et al.*, *Nat. Mater.* **3**, 632–637 (2004).

24. S. M. Sze, Ng, *Physics of semiconductor devices* (Wiley-Interscience, Hoboken, N.J., 2007).

25. A. D. McLachlan, F. P. Meyer, *Appl. Opt.* **26**, 1728–1731 (1987).

26. H. Berthou, C. K. Jörgensen, *Opt. Lett.* **15**, 1100–1102 (1990).

**Acknowledgment**s

The authors would like to acknowledge Prof. Peter Wurfel and Prof. Eli Yablonovitch for their support, fruitful discussions and valuable insights. The authors would also like to thank Assist Prof. Avi Niv for the constructive brainstorming.

**Funding**: This report was partially supported by the by the Russell Berrie Nanotechnology Institute (RBNI), and the Grand Technion Energy Program (GTEP) and is part of The Leona M. and Harry B. Helmsley Charitable Trust reports on Alternative Energy series of the Technion and the Weizmann Institute of Science. We also would like to acknowledge partial support by the Focal Technology Area on Nanophotonics for Detection. A. Manor thanks the Adams Fellowship program for the financial support.

# Optical Refrigeration for Ultra-efficient Photovoltaics

# Supplementary Material

## A. The detailed balance calculation:

In steady-state, under QE of unity, both the rate and energy conditions have to be satisfied. Specifically, for the absorber: The rate of the absorbed photons from the sun and the PV is balanced by its photoluminescence, which by the generalized Planck's equation includes also thermal emission. The energy balance for the absorber is equal to the rate balance with the multiplication by the photon energy quanta, $\hbar\omega$, in the rate equation. For the PV: The rate of absorbed photons from the intermediate absorber is balanced by the current extraction and its spontaneous emission. The energy conservation at the PV does not apply since heat is evacuated in order to maintain the PV at 300K.

**The rate equations are:**

*Absorber (1):*

$$\underbrace{\Omega_s \cdot \int_{E_{g1}}^{\infty} R_{sun}(T_s)dE}_{from\_sun} + \underbrace{\pi \cdot \int_{E_{g2}}^{\infty} R_{PV}(300,V)dE}_{from\_PV} = \underbrace{\pi \cdot \int_{E_{g2}}^{\infty} R_{abs}(T,\mu)dE}_{to\_PV} + \underbrace{\Omega_1 \cdot \int_{E_{g1}}^{\infty} R_{abs}(T,\mu)dE}_{to\_sky}$$

*PV (2):*

$$\underbrace{\pi \cdot \int_{E_{g2}}^{\infty} R_{abs}(T,\mu)dE}_{from\_abs} = \underbrace{\pi \cdot \int_{E_{g2}}^{\infty} R_{PV}(300,V)dE}_{to\_PV} + \frac{I}{e}$$

Where $\Omega$ is the solid angle, $I$ is the electron current and $R$ is the rate given by the generalized Planck law:

$$R(\hbar\omega,T,\mu) = \varepsilon(\hbar\omega) \cdot \frac{(\hbar\omega)^2}{4\pi^2\hbar^3 c^2} \frac{1}{e^{\frac{\hbar\omega-\mu}{KT}}-1} \cong \varepsilon(\hbar\omega) \cdot \frac{(\hbar\omega)^2}{4\pi^2\hbar^3 c^2} e^{\frac{-\hbar\omega}{K_b T}} e^{\frac{\mu}{K_b T}}$$

Where $\varepsilon$ is the emissivity, $K_b$ is Boltzmann's constant, $T$ is the temperature, $\mu$ is the chemical-potential, $c$ is the speed of light in vacuum and $\hbar\omega$ is the photon's energy quanta. The emissivity is taken as a step function which corresponds to the material's bad-gap, i. e zero below $E_g$ and unity above it. The energy equations are:

*Absorber (1):*

$$\overbrace{\Omega_s \cdot \int_{E_{g1}}^{\infty} E_{sun}(T_s)dE}^{from\_sun} + \overbrace{\pi \cdot \int_{E_{g2}}^{\infty} E_{PV}(300,V)dE}^{from\_PV} = \overbrace{\pi \cdot \int_{E_{g2}}^{\infty} E_{abs}(T,\mu)dE}^{to\_PV} + \overbrace{\Omega_1 \cdot \int_{E_{g1}}^{\infty} E_{abs}(T,\mu)dE}^{to\_sky}$$

*PV (2):*

$$\overbrace{\pi \cdot \int_{E_{g2}}^{\infty} E_{abs}(T,\mu)dE}^{from\_abs} = \overbrace{\pi \cdot \int_{E_{g2}}^{\infty} E_{PV}(300,V)dE}^{to\_PV} + \frac{I}{e}$$

And:

$$E(\hbar\omega,T,\mu) = \hbar\omega \cdot R(\hbar\omega,T,\mu)$$

The two equations are simultaneously solved. The solution is unique by fixing the only free parameter in the system; the PV voltage. In cases of thermal regime the solution sets at negative values for the chemical-potential ($\mu<0$). In this case the rate equation is discarded and the simulation is solved only for energy conservation with $\mu=0$. It is important to note that the equality of the integrated rates requires that the number of excitations change only by absorption and emission of photons. That excludes non-radiative transitions by which excitations disappear, but it also excludes reactions among the excitations, e.g. one high energy e-h pair gives two lower energy e-h pairs. Such a mechanism would be Auger recombination and impact ionization. These are mechanisms which are fundamental and cannot, in principle, be excluded for semiconductors. However for ionic fluorescence materials, such as rare earths, the excited electron is localized and such non-radiative mechanisms are negligible.

### B. Definition of the photon recycling angle and comparison to solar concentration :

In conventional solar concentration, as the concentration factor increases the device "sees" the sun from larger solid angle. For example, at maximal solar concentration, of 45200, the sun covers the entire hemispherical solid angle of $2\pi$. Normalizing for flat panel, which sees the projected solid angle per area, leads to average correction factor of 1/2. That is the maximal solid angle is $\Omega_{sun}=\pi$ *Srad*.

Because increasing the solid angle at which solar radiation reaches the device inherently allows PL to escape, we first use the detailed balance to optimize $\Omega_1$. We find that the optimal configuration is a non-concentrated solar illumination with full photon-recycling. That is, the acceptance solid angle of the sun equals its viewing angle without any use of concentrating lens. In this case $\Omega_1=\Omega_{sun}=6.94 \cdot 10^{-5}$ *Srad*, while the rest of the hemisphere angles are reflected back to the device by the reflective surface. This optimum is because the recycled photons already thermalized and are therefore considered as a low entropy source ideal for optical cooling. The equivalent concentration factor, $X_{eq}$, to the ratio of photon recycling is defined by:

$$X_{eq} = \frac{\pi}{\Omega_1}, \quad X_{max} = \frac{\pi}{\Omega_{sun}} \approx 45200,$$
$$X_{eq} = 5000 \rightarrow \Omega_1 = 9.05 \cdot \Omega_{sun}$$

### C. The Fluorescence Intensity Ratio Thermometry (FIR) method:

Using the FIR method, temperature can be measured by recording the power spectrum intensity ratio, of two adjacent optical transitions of a luminescent sample. Under Boltzmann population statistics, the ratio is defined by:

$$R = C \cdot e^{(-\frac{\Delta E}{K_b T})}$$

Where $\Delta E$ is the energy difference between the two levels. The constant, $C$, is related to the density of states and the matrix element for optical transitions between the states.

Once $C$ is known the relation can be used to determine the sample's temperature during the experiment.

In order to find $C$, we use a calibrated micro-furnace, which consists of a heated Quartz cylinder. From the same Neodymium ($Nd^{+3}$) doped Silica fiber as in the experiment, we cut a 5mm long fiber and connect (fuse) it, from both sides, to a long passive optical fiber (Figure S1A). The 5mm active fiber section is placed at the center of the furnace, while the fiber is pumped by 532nm laser from one side, and light is collected from its other side. A spectrum analyzer is used to measure the ratio between the 900nm and 820nm peaks as function of the furnace temperature. Figure S1B depicts the experimental intensity-ratio Vs. temperature, and the fitted curve. The constant is found to be $C=2.6$. The inset shows the evolution of the peak ratio shift with temperature increase.

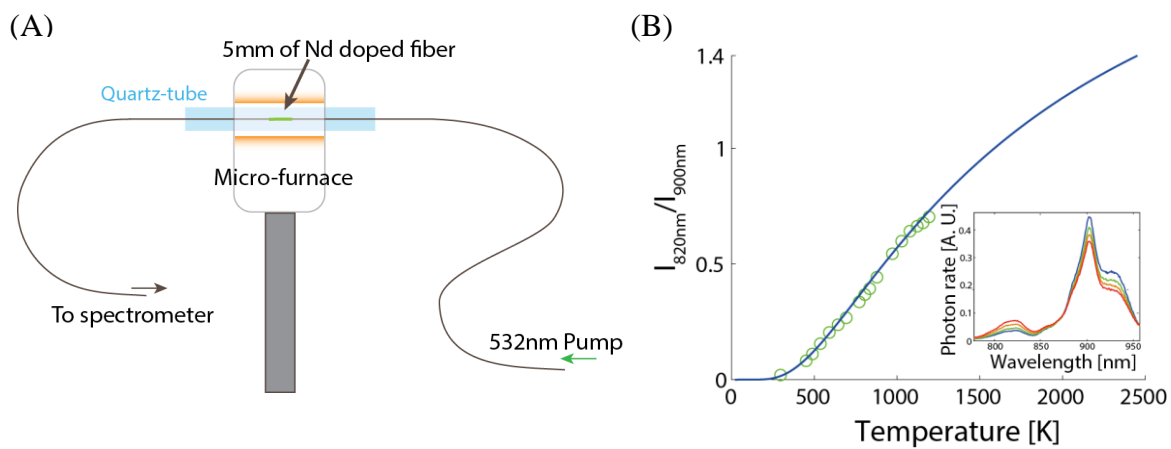

Figure S1. (A) The FIR measurement experimental setup (B) The intensity ratio-temperature plot.